\documentclass[11pt,a4paper,floatfix]{revtex4}
\usepackage[utf8x]{inputenc}
\usepackage[english]{babel}
\usepackage{amssymb,amsmath,amsfonts}
\usepackage{graphicx}
\usepackage{hyperref}
\usepackage[usenames]{color}

\newcommand{\bkbo}{Ba$_{1-x}$K$_x$BiO$_3$ }

\begin{document}

\title{Correlation effects and phonon modes softening with doping in Ba$_{1-x}$K$_x$BiO$_3$}
\author{Dm.\,M. Korotin}
\affiliation{Institute of Metal Physics, S.~Kovalevskoy St. 18, 620990 Yekaterinburg, Russia}
\author{D.~Novoselov}
\affiliation{Institute of Metal Physics, S.~Kovalevskoy St. 18, 620990 Yekaterinburg, Russia}
\affiliation{Ural Federal University, Mira St., 19, 620002 Yekaterinburg, Russia}
\author{V.~I.~Anisimov}
\affiliation{Institute of Metal Physics, S.~Kovalevskoy St. 18, 620990 Yekaterinburg, Russia}
\affiliation{Ural Federal University, Mira St., 19, 620002 Yekaterinburg, Russia}

\begin{abstract}
Monoclinic crystal structure of the undoped BaBiO$_3$ can be described as a cubic perovskite  distorted by a frozen breathing and tilting phonon modes of BiO$_6$ octahedra.
The phonon mode softening is experimentally observed~[M. Braden {\em et al.}, Europhysics Letters (EPL) 34, 531 (1996)] in \bkbo through potassium doping followed by a transition into an ideal cubic perovskite structure at $x=0.37$ close to the appearance of superconductivity.
In our previous paper~[D. Korotin {\em et al.}, Journal of Physics: Condensed Matter 24, 415603 (2012)] we demonstrated that it is necessary to take into account correlation effects by DFT+U method  in Wannier functions basis to obtain a good agreement between the calculated and  experimental values of crystal structure distortion and  energy gap in BaBiO$_3$.
In the present work with the same method we calculated the breathing mode phonon frequencies as a function of potassium doping level in Ba$_{1-x}$K$_x$BiO$_3$. 
Obtained frequencies are in a good agreement with experimental values and the breathing mode softening with doping effect is reproduced while calculations without consideration of correlation effects failed to do so.
We shown that the cubic crystal structure becomes stable at $x=0.30$ in agreement with the experimental transition to cubic perovskite at $x=0.37$.
The possible connections between the correlation effects, phonon mode softening, and superconductivity in \bkbo are discussed.

\end{abstract}

\maketitle

\section{Introduction}
The mechanism of the high-temperature superconductivity in \bkbo is still the open question. 
This material has a rather high critical temperature $T_c \approx 30 K (x \approx 0.4)$ for a system without a layered structure in contrast to CuO planes in cuprates~\cite{Mattheiss1988, Hinks1988, Cava1988, Batlogg1988} and Fe-As planes in pnictides and without any {\em d-} or {\em f-}elements at all.
Furthermore, the absence of magnetic fluctuations that are  important for pairing in copper oxides and iron based superconductors~\cite{Dagotto1994, Chen2008} assumes different than in cuprates and pnictides mechanism of superconductivity.

There are various and partly contradictory data about the relevance of traditional electron-phonon mechanism of SC in this material. 
Accumulated results of numerous researches do not allow to determine unambiguously the strength of electron-phonon coupling in Ba$_{1-x}$K$_x$BiO$_3$.
For example, large oxygen isotope effect $\alpha = 0.4$ measured at the replacement $^{18}$O by $^{16}$O in Ba$_{0.625}$K$_{0.375}$BiO$_3$ compound indicates  the conventional phonon-mediated superconductivity mechanism~\cite{Hinks1988a}.
At the same time Batlogg~{\em et al.} ~\cite{Batlogg1988} obtained the value $\alpha = 0.21 \pm 0.03$ 
%from analysis of the density of states 
and concluded that \bkbo is a weak-coupling HTSC.
Marsiglio and colleagues~\cite{Marsiglio1996} came to the same conclusion getting the value of the dimensionless electron-phonon coupling constant $\lambda \approx 0.2$. 
On the other hand, the results of the inelastic neutron scattering experiments carried out by Braden {\em et al.}~\cite{Braden1995a} showed the substantial frequency shift between the lightly doped Ba$_{0.98}$K$_{0.02}$BiO$_3$ and Ba$_{0.6}$K$_{0.4}$BiO$_3$. That frequency shift can be interpreted as a result of the strong electron-phonon coupling.
In work~\cite{Graebner1989} devoted to the measurements of the heat capacity of single crystals of Ba$_{0.6}$K$_{0.4}$BiO$_3$ in the vicinity of superconducting transition temperature the value $\lambda \approx 0.35$ was obtained.
High-resolution inelastic X-ray scattering results for $x = 0, 0.30, 0.37$, and 0.52 on the \bkbo crystal presented in paper~\cite{Khosroabadi2011} point to existence of anomalous softening at $x = 0.37$ and 0.52 for volume and one-dimensional breathing modes which also agrees with the results published in the works~\cite{McCarty1989, Nishio2003}.
Sharifi and colleagues~\cite{Sharifi1991} from the results of tunneling measurements experiment have obtained the values $\lambda \approx 0.2$ and $2\Delta / kT_c = 3.5$ and argued that the electron-phonon interaction in \bkbo has the weak-coupling character.
The authors note that these results relate to the interaction of electrons with acoustic phonon modes only in contrast to optical branch breathing phonon mode considered in the present work.
However, the evidence of the importance of an electron-phonon interaction in \bkbo has been obtained from several tunneling measurements experiments~\cite{Huang1990, Samuely1993, Tralshawala1995}.

The results of theoretical investigations of the superconductivity related properties in \bkbo also lead to controversial conclusions on the role of the electron-phonon interaction.
The authors of~\cite{Shirai1990a} using the band structure calculations have shown that for $x > 0.3$ the $\lambda$ value exceeds 1 thus placing 
 \bkbo  to the class of strong-coupling superconductors.
In addition, the authors concluded that $\lambda$ sharply decreases with increasing $x$ in consequence of significant renormalization of the breathing modes playing an important role in the high-$T_c$ superconductivity phenomenon in Ba$_{1-x}$K$_x$BiO$_3$.
Hamada {\em et al.}~\cite{Hamada1989} have calculated the electron-phonon coupling constant value $\lambda = 3$ within LDA approximation.
Investigating the optical properties of \bkbo Nourafkan~{\em et al.}~\cite{Nourafkan2012} have employed dynamical mean field theory and found the strong dependence of the effective electron-phonon interaction on the doping with the largest value of $\lambda \geq 4$ at $x = 0.4$.

On the other hand, Liechtenstein and coworkers~\cite{Liechtenstein1991} applying the LDA method have obtained the estimate of the value $\lambda_{b} \approx 0.3$ for the breathing mode for $x$ from 0.37 to 0.5.
The work of  Kunc and Zeyher~\cite{Kunc1994} contains a rough estimate of $\lambda \approx 0.5$ obtained in LDA calculations.
Mergali and Savrasov~\cite{Meregalli1998} applying LDA and LDA+U method have obtained the value $\lambda = 0.29$ that is too small to explain the electron-phonon nature of superconductivity at $T_c = 30 K$ in Ba$_{0.6}$K$_{0.4}$BiO$_3$ compound.

\begin{figure}[tbp!]
\centerline{\includegraphics[width=0.5\columnwidth,clip]{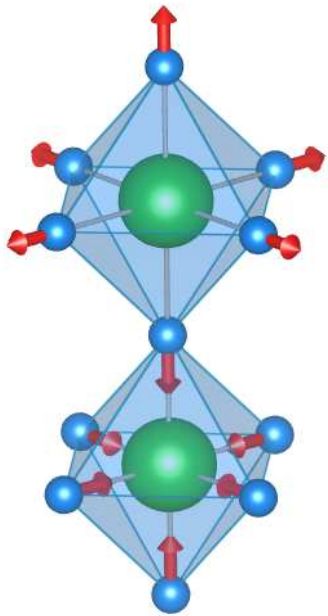}}
\caption{(color online) Schematic view of the BiO$_6$ octahedra breathing distortions in \bkbo.
Blue spheres denote oxygen ions, green spheres correspond to bismuth ions, red arrows depict the directions of displacements.}
\label{fig:structure}
\end{figure}

% It was concluded that the breathing distortions is significantly underestimated (or practically absent) in local density approximation.
Hence the questions of reliable description of the electron-phonon coupling strength in \bkbo and whether the conventional phonon mechanism of superconductivity is essential in this system or not still remain unclear. The phonon mode softening effect that is experimentally observed~\cite{Braden1996} in \bkbo through potassium doping can be connected with superconductivity appearance in this material if the corresponding phonon modes are strongly coupled with electrons at the Fermi level~\cite{Vonsovsky1982, Phillips1972}. 
%{\em kakie nibud ssylki na svyz smyagchenia b sverhprovodimosti????} 
The softening effect appears with potassium doping for the high-energy optical phonon mode~\cite{Mattheiss1988, Shirai1990a,Loong1989} corresponding to a breathing distortion of BiO$_6$ octahedra, see Fig.~\ref{fig:structure}. In the present work we calculated  the breathing mode phonon frequencies as a function of potassium doping level in Ba$_{1-x}$K$_x$BiO$_3$. 

In the previous paper~\cite{Korotin2012} we have shown that it is necessary to take into account correlation effects in the framework of GGA+U approximation in Wannier functions basis in order to reproduce  successfully features of crystal and electronic structures of BaBiO$_3$. The same method was applied to the problem of phonon softening in  \bkbo in the present work and allowed us reproduce successfully the experimentally observed effect. In contrast, standard DFT calculations failed to produce such a result. We also estimated the effect of correlations on increasing the strength of electron-phonon coupling in \bkbo in comparison with standard DFT results.

\section{Calculation method}

Our calculations were performed with plane-waves pseudopotential approach implemented in Quantum-ESPRESSO package~\cite{Giannozzi2009}. We used pseudopotentials in the ultrasoft form with PBE functional. The plane waves energy cutoff was set to 50~Ry and the charge density cutoff equals to 400~Ry. Integrations in reciprocal space were performed using (16,16,16) Monkhorst-Pack {\bf k}-point grid in the full Brillouin zone. 
The total energy convergence limit was set to 10$^{-9}$~Ry to obtain an accurate total energy dependence.

Several modifications of the code were implemented for DFT+U method in Wannier functions basis as described in~\cite{Korotin2008a}. In particular the total energy is computed as
\begin{equation}
	E^{tot} = E^{DFT} + E^{U} - E^{DC},
\end{equation}
where $E^{DFT}$ is the total energy from a standard GGA calculation, $E^{U} = \frac{1}{2}\sum_{m\neq m\prime}Un_mn_m\prime$ is the Coulomb interaction correction and $E^{DC} = \frac{1}{2}Un(n-1)$ is the double counting correction.  We define $n_m$ as $m$th Wannier function occupation number and $n=\sum_m n_m$ as the total occupancy of WFs. 

The breathing distortions of BiO$_6$ octaredra were treated in cubic $Fm\bar3m$ cell with two formula units. 
Potassium doping was simulated with rigid band approximation.
The pseudocubic lattice parameter is defined as $a_p = (4.3548-0.1743x)$, where $x$ -- is the potassium concentration~\cite{PhysRevB.41.4126}. 

WFs were generated by projection of two Bi-$s$ atomic orbitals onto subspace defined by the two energy bands near the Fermi level. 
The Hubbard $U$ value for BaBiO$_3$ equals 0.7~eV~\cite{Korotin2012}. The same value is used in our calculations.

The breathing phonon frequency at $R$ point was computed within frozen-phonon approach. The phonon frequency is proportional to the second derivative of the total energy over atomic displacement. The total energy vs breathing distortion dependence was interpolated with a parabola. The interpolation was performed using an ordinary least squares method in a region near the total energy minima where harmonic approximation is reliable. Then the second order derivative was taken analytically. 

\section{Results and discussion}

%%%%%%%%%%%%%%%%%%%%%%%%%%%%%%%%%%%%%%%%%%%%%%%%%%%%
\begin{figure}[tbp!]
\centerline{\includegraphics[width=0.7\columnwidth,clip]{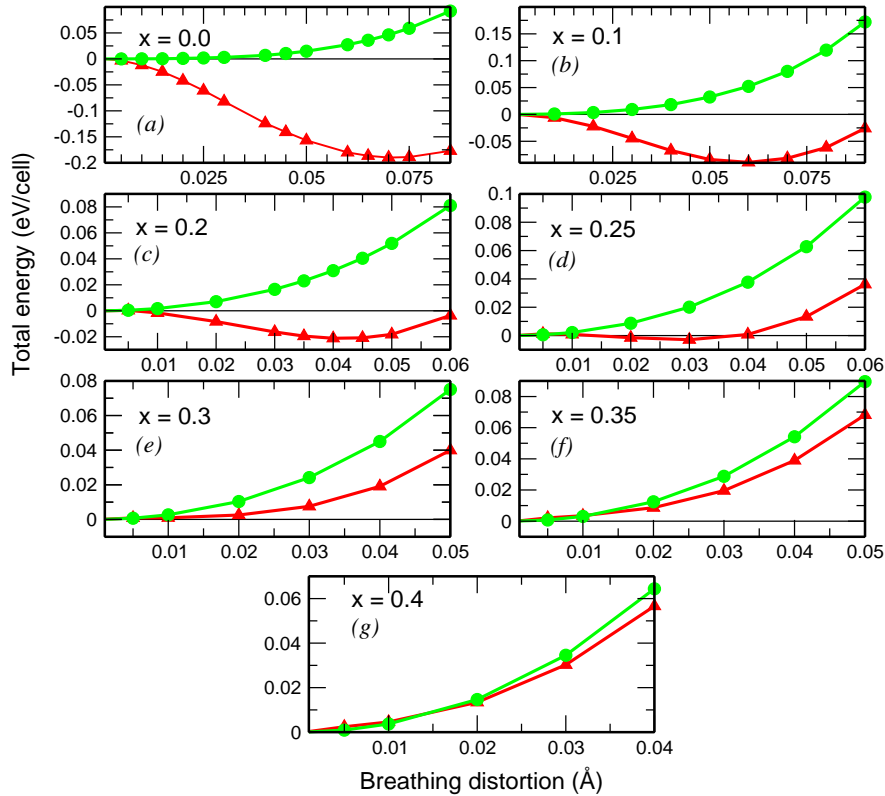}}
\caption{(color online) Total energy of \bkbo as a function of BiO$_6$ octahedra distortion calculated for various potassium doping. Green curve with circles corresponds to GGA calculation, red with triangles curve corresponds to GGA+U in Wannier functions basis calculation.}
\label{fig:toten}
\end{figure}
%%%%%%%%%%%%%%%%%%%%%%%%%%%%%%%%%%%%%%%%%%%%%%%%%%%%

In Fig.~\ref{fig:toten} the total energy of \bkbo is shown as a function of BiO$_6$ octahedra breathing distortion. 
The calculations were done within GGA and GGA+U in Wannier functions basis approaches.
The GGA and GGA+U curves were shifted in such a way that the total energy values for undistorted structure were set to zero. 

Let us first discuss the results obtained by standard GGA without correlations taken into account.
In GGA the total energy dependence curves (green lines on Fig.~\ref{fig:toten}) are qualitatively the same for all potassium doping levels. 
The total energy has a parabolic form with a minimum at zero distortion that means a stable cubic perovskite crystal structure for every potassium concentration. 
This result is in disagreement with the experimentally observed monoclinic structure for BaBiO$_3$~\cite{Cox1976969} and orthorhombic structure for $x=0.1 .. 0.35$~\cite{PhysRevB.41.4126}. 
As it was shown previously for undoped compound BaBiO$_3$~\cite{Korotin2012} such a result is caused by the neglect of Coulomb correlations between partially filled states having mixed Bi-{\em 6s} and O-{\em 2p} orbitals origin. 
In the paper~\cite{Korotin2012}
Coulomb correlations were taken into account using  GGA+U method that has allowed to reproduce successfully combined effect of breathing and tilting distortion of the BiO$_6$ octahedra in a monoclinic structure.

Correlation effects taken into account
in the GGA+U calculations changed the calculated results (red lines in Fig.~\ref{fig:toten}) qualitatively comparing with the GGA ones. 
The cubic crystal structure is now unstable against breathing distortion for pure BaBiO$_3$, see Fig.~\ref{fig:toten} (a), in the agreement with results of ~\cite{Korotin2012}.
The total energy has a minimum at 0.075$\AA$ that is in a good agreement with the experimental value 0.085$\AA$~\cite{Cox1976969}. 
One should take into account that in the present paper tilting of the BiO$_6$ octahedra existing in a real monoclinic structure was neglected while in the previous paper ~\cite{Korotin2012} for undoped BaBiO$_3$ tilting was treated on the same footing as breathing distortion. 
 
The depth of the total energy minimum decreases with increaseof the doping level. 
The theoretical structural transition to an ideal cubic perovskite cell happens for Ba$_{0.7}$K$_{0.3}$BiO$_3$ (Fig.~\ref{fig:toten} (e)) where minimum for final distortion value disappears and energy is minimal for the undistorted cubic structure. 
The calculated value of critical potassium concentration $x = 0.30$ where cubic structure becomes stable is in a good agreement with the experimental estimations of orthorhombic to cubic structure transition at 0.37~\cite{PhysRevB.41.4126}. 
The slight difference between the experimental value and our estimations could arise from the neglect of the tilting distortion. 
In the cubic phase (for $x=0.3..0.45$) a curvature of the total energy dependence increases with the increase of doping.

%%%%%%%%%%%%%%%%%%%%%%%%%%%%%%%%%%%%%%%%%%%%%%%%%%%%
\begin{figure}[tbp!]
\centerline{\includegraphics[width=0.5\columnwidth,clip]{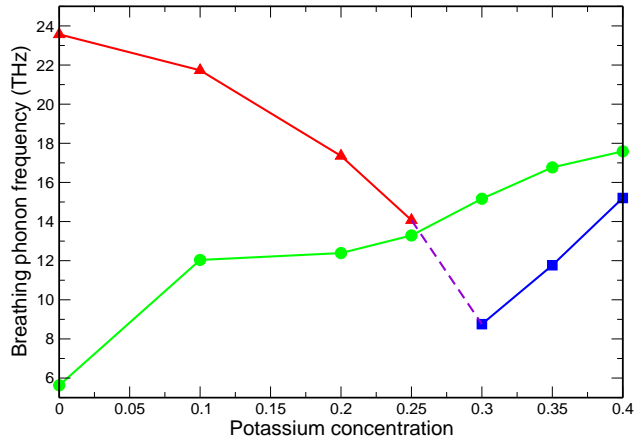}}

\caption{(color online) Breathing phonon mode frequency as a function of potassium doping calculated within GGA (green curve with circles) and GGA+U (red curve with triangles for distorted structure and blue squares for the ideal cubic structure).} 
\label{fig:omega}
\end{figure}
%%%%%%%%%%%%%%%%%%%%%%%%%%%%%%%%%%%%%%%%%%%%%%%%%%%%

From the data shown in Fig.~\ref{fig:toten} the second derivative of the total energy that is proportional to the breathing phonon frequency was computed as a function of doping $x$. The result is shown in Fig.~\ref{fig:omega}.
%There are three frequency vs doping dependencies computed for BiO$_6$ breathing mode. 
The results of the GGA calculations are shown by the green curve with circles. An ideal cubic perovskite structure is stable in GGA for the whole range of doping levels.
The breathing phonon mode frequency increases with the potassium concentration increase from 5.6~THz in pure BaBiO$_3$ to 17.6~Thz in Ba$_{0.6}$K$_{0.4}$BiO$_3$ thus demonstrating effect of hardening phonon mode instead of experimentally observed softening. There is no any evidence of structural phase transition. 

There are two curves for frequency vs doping dependence for the breathing mode calculated in GGA+U marked by red and blue line in Fig.~\ref{fig:omega}. 
The reason for that is the fact that there are two different stable crystal structures in GGA+U (according to the total energy calculations shown in Fig.~\ref{fig:toten}). 
For $x=0..0.25$ the breathing distortion of BiO$_6$ octahedra exists in a ground state of \bkbo so the phonons correspond to deviation of atoms from equilibrium positions in distorted crystal structure.
The corresponding breathing mode frequency is shown in Fig.~\ref{fig:omega} by red curve with triangles. 
The frequency decreases with doping from 23.6~THz for BaBiO$_3$ to 14.1~THz for Ba$_{0.75}$K$_{0.25}$BiO$_3$. 
In agreement with the experimental data~\cite{Braden1996} the breathing mode softening is reproduced in our GGA+U calculation.

The theoretical structural transition occurs at potassium concentration equal to $x=0.30$. 
Starting from that concentration the ground state crystal structure is the ideal cubic structure and phonons correspond to the deviation of atoms from equilibrium positions in cubic perovskite. 
The breathing phonon frequency for the ideal cubic structure is shown in Fig.~\ref{fig:omega} by blue line with squares. The frequency increases with doping from 8.8~THz for Ba$_{0.7}$K$_{0.3}$BiO$_3$ to 15.2~THz for Ba$_{0.6}$K$_{0.4}$BiO$_3$. 
It should be noted that GGA and GGA+U frequencies lie close to each other at the point $x=0.4$. 
In Fig.~\ref{fig:toten} (g) the green and red curves corresponding to the GGA and GGA+U results for total energy as a function of distortion are also very close to each other. It means that correlations become unimportant for ``overdoped" case $x\ge 0.4$.

The obtained breathing mode frequency in the cubic phase is in a good agreement with experimental data. 
For example, for $x=0.4$ calculated value 15.2~THz is very close to the experimental value 15.0~THz obtained by Braden {\em et al.}~\cite{Braden1996}. 
The agreement between calculated (23.6~THz) and available experimental data (17.05~THz~\cite{Tajima1992} and 16.9~THz~\cite{Braden1996}) for undoped phase is not so good, but reasonable. 
%The difference is caused by model, not real monoclinic crystal structure for $x < 0.3$ in our calculations. 
For the intermediate doping level $x=0.2$ the agreement (17.35~THz vs 17.05~\cite{McCarty1989}) is good. 

The phonon softening is observed for GGA+U curves in Fig.~\ref{fig:omega} when system approaches to the crystal structure transition. The effect is seen as concentration increases  $0<x<0.3$ destabilizing distorted crystal structure before transition to cubic perovskite, as well as for decreasing concentration values in the range $0.3<x<0.4$ in its turn destabilizing cubic structure as the system approaches the distorted crystal structure.
%{\em zdes' nuzhny umnye slova o svyazi smyagchenia phononov vblizi strukturnogo perehoda i sverhprovodimosti}
As it is well known that a structural instability near $T_c$ accompanied by a softening of phonon modes often causes a superconductivity transition~\cite{Vonsovsky1982, Phillips1972, Testardi1974, Collings1972}.

%We were unable to reproduce the semiconductor-metal transition because of rigid-band approximation used for the potassium doping simulation. 
%Within the rigid-band approximation the \bkbo become metallic for any non-zero doping level. 
%The slight change in the carriers number shifts the Fermi level into the conduction band. 
%It was shown~\cite{Franchini2009} that the semiconducting behavior in doped BaBiO$_3$ is a polaronic effect.
%We do not consider polarons in the present calculations and don't reproduce semiconducting electronic structure for the doped compound. 
%However the phonon softening effect was reproduced successfully. 
%One can conclude that the described softening is related to the structural transition. 

In the present work we demonstrated that correlation effects taken into account in GGA+U method could qualitatively change calculated phonon frequency as a function of potassium concentration in  Ba$_{1-x}$K$_x$BiO$_3$. In order to understand the problem of superconductivity mechanism in this material it would be necessary to calculate the full phonon spectra with the corresponding electron-phonon coupling constants $\lambda$. That is a computationally hard task with correlation effects taken into account and we plan to perform it in a separate work. However, we could try to estimate roughly the influence of the correlation effects on the $\lambda$ for breathing phonon mode considered here.

In paper~\cite{Meregalli1998} it was shown that $\lambda$ for the breathing phonon mode is an order of magnitude larger than $\lambda$ for the other modes so that total strength of electron-phonon coupling will be defined predominantly by this mode. We analyze only one ${\bf q}$-vector corresponding to the $R$ high-symmetry point in the Brillouin zone.
The {\em el-ph} coupling constant for one mode and wave-vector ${\bf q}$ is defined as~\cite{Martin2004}:
\begin{equation}
	\label{eq:lambda}
	\lambda = \frac{2}{N(E_f)}\frac{1}{\omega_{\bf q}} \sum_{ij{\bf k}} {|g_{i{\bf k};j{\bf k+q}} |}^2 \delta(\epsilon_{i{\bf k}}) \delta(\epsilon_{j{\bf k+q}} - \epsilon_{i{\bf k}} - \omega_{\bf q}),
\end{equation}
where $\omega_{{\bf q}}$ is the mode frequency, $N(E_f)$ is the electronic density of states at the Fermi level, $\epsilon_{i{\bf k}}$ is the energy of electron band $i$ with the wave-vector ${\bf k}$, $g_{i{\bf k};j{\bf k+q}}$ is the electron-phonon matrix element:
\begin{equation}
	\label{eq:g}
	g_{i{\bf k};j{\bf k+q}} = \frac{1}{\sqrt{2M\omega_{\bf q}}} \langle i{\bf k} | \frac{\partial V}{\partial{\bf u}_{b}} | j{\bf k+q} \rangle,
\end{equation}
where $M$ is mode dependent reduced mass, and $\Delta V_{\bf q}$ is the change in electronic potential $V$ due to atomic displacements corresponding to the phonon. Finally the electron-phonon coupling constant is proportional to
\begin{equation}
\label{eq:lambda2}
	\lambda \sim \frac{{|\langle i{\bf k} | \frac{\partial V}{\partial{\bf u}_{b}} | j{\bf k+q} \rangle|}^2}{\omega^2_{\bf q}}.
\end{equation}

Lets consider the $x=0.30$ point in Fig.~\ref{fig:omega} that corresponds to theoretical structural transition.
The breathing mode frequency obtained within the GGA+U calculation equals 8.8~THz that is 1.7 times smaller than in GGA (15.2~THz).
Thus electronic correlations result in a 2.9 times decrease of $\omega^2$ term in denominator of Eq.~\ref{eq:lambda2}.

To estimate the change in electronic potential due to atomic displacements, we used our results of bands structure calculations for pure BaBiO$_3$ presented in Fig.~2 of our previous paper~\cite{Korotin2012}. 
In order to estimate
$\langle i{\bf k} | \frac{\partial V}{\partial{\bf u}_{b}} | j{\bf k+q} \rangle$ 
element of electron-phonon matrix, we consider a change of one-electron energies due to the frozen breathing phonon distortions:
\begin{equation}
	\delta \epsilon_{i,\bf k} = \langle i{\bf k} | \frac{\partial V}{\partial{\bf u}_{b}} \cdot u_b | i{\bf k} \rangle,
\end{equation}
where ${u}_{b}$ is the frozen breathing phonon distortion. 
The change in numerator of Eq.~\ref{eq:lambda2} we evaluate as:
\begin{equation}
\label{rat}
	\frac{\langle i{\bf k} | \partial V^{GGA+U}/ \partial {\bf u} | j{\bf k+q} \rangle}{\langle i{\bf k} | \partial V^{GGA}/ \partial {\bf u} | j{\bf k+q} \rangle} = \frac{\delta \epsilon^{GGA+U}_{i}}{\delta \epsilon^{GGA}_{i}}, 
\end{equation}
where $\epsilon^{GGA+U}_{i}$ is the one-electron energy of the lowest empty band at $A$ point.  
From analysis of electronic band structure of the ideal cubic and distorted BaBiO$_3$ presented Figs.~2({\em a})-({\em a}) in~\cite{Korotin2012} we obtained $\delta \epsilon^{GGA} = 0.75~eV$ and $\delta \epsilon^{GGA+U} = 1.1~eV$ that gives us a ratio  \eqref{rat} value $\approx$ 1.5. 
That results in a 2.25 times increase of the numerator of Eq.~\ref{eq:lambda2}. 
Hence, the total increase of the fraction value in Eq.~\ref{eq:lambda2} due to correlation effects is 6.5 times. 

The previous calculations of $\lambda$ within LDA by Liechtenstein {\em et al.}~\cite{Liechtenstein1991} 
($\lambda_{b} \approx 0.3$) and by Kunc and Zeyher~\cite{Kunc1994} ($\lambda \approx 0.5$) 
did not allow to state the strong coupling regime in Ba$_{1-x}$K$_x$BiO$_3$.
Our estimates show that value of electron-phonon coupling will be $\lambda > 1$ that assumes a strong-coupling electron-phonon interaction in this compound.

\section{Conclusion}
The breathing phonon softening with potassium doping in \bkbo was obtained as a result of {\em ab initio} calculation by GGA+U method for the first time. A transition from distorted crystal structure to the ideal cubic perovskite was theoretically found at $x\approx 0.3$ in agreement with the experimental data.
Coulomb correlations are found to be essential for the proper description of crystal structure and vibrations properties of the compound. Standard GGA calculations result in a phonon hardening instead of softening. The estimations of  electron-phonon coupling enhancement with correlation effects taken into account put \bkbo in 
strong electron-phonon coupling regime.
%But for the straightforward statement it is necessary to compute full phonon spectra for various doping levels and estimate the electron-phonon coupling constant value. 
%Such calculations are being performed and the results will allow to close the question about the coupling strength in \bkbo. 
%At the moment we want to accentuate the key physical property for a successive crystal structure distortions description and the suitable computational method.

\section{Acknowledgments}
This work was performed using ``Uran'' supercomputer of IMM UB RAS and supported by the Russian Foundation for Basic Research (projects Nos. 12-02-31257, 13-02-00050, 12-02-91371-CT), by the Deutsche Forschungsgemeinschaft through TRR 80 and FOR 1346, by the Fund of the President of the Russian Federation for the Support of Scientific Schools NSH-6172.2012.2, and by The Ministry of education and science of Russian Federation project 14.A18.21.0076. 
D. Novoselov is grateful to the Dynasty Foundation. 

\bibliography{BaKBiO3}

\end{document}